\documentstyle[12pt,epsf]{article}

\newcommand{\e}{{\rm  e}}

\newcommand{\beq}{ \begin{eqnarray} }
\newcommand{\eeq}{ \end{eqnarray} }
\newcommand{\beqstar}{ \begin{eqnarray*} }
\newcommand{\eeqstar}{ \end{eqnarray*} }
\newcommand{\gsim}{ \mathop{}_{\textstyle \sim}^{\textstyle >} }

\begin{document}
\baselineskip 0.7cm

\begin{titlepage}

\begin{center}

\hfill KEK-TH-575, UT-815\\
\hfill OU-HET-294\\
\hfill hep-ph/yymmdd\\
\hfill \today

  {\large  
	Enhancement of 
	$\mu\rightarrow \e \gamma$ \\
	in the supersymmetric SU(5) GUT
	at large $\tan\beta$}
  \vskip 0.5in {\large
    J.~Hisano$^{(a)}$,
    Daisuke~Nomura$^{(b)}$,
    Yasuhiro~Okada$^{(a)}$,
    Yasuhiro~Shimizu$^{(a)}$, and 
    Minoru~Tanaka$^{(c)}$ }
\vskip 0.4cm 
{\it 
(a) Theory Group, KEK, Oho 1-1, Tsukuba, Ibaraki 305-0801, Japan
}
\\
{\it 
(b) Department of Physics, University of Tokyo, Tokyo 113-0033, Japan
}
\\
{\it 
(c) Department of Physics, Graduate School of Science, 
	Osaka University, Toyonaka, Osaka 560-0043, Japan
}
\vskip 0.5in

\abstract {
Branching ratio of $\mu \rightarrow \e \gamma$ is evaluated in 
the supersymmetric SU(5) grand unified theory taking account of higher 
dimensional operators, 
which reproduce the realistic quark and lepton masses and the mixing parameters.
It is shown by numerical calculation that at large $\tan\beta$ the branching
ratio is enhanced by a few 
orders of magnitude since the flavor mixings for both the left- 
and the right-handed sleptons are induced by the effect of the higher 
dimensional operators.
}
\end{center}
\end{titlepage}

\setcounter{footnote}{0}

The supersymmetric grand unified theory (SUSY GUT),
unifying three gauge groups in the Standard Model (SM), is one of the most interesting 
models from experimental and theoretical points of view. This model explains
the electric-charge quantization automatically. The weak mixing angle 
predicted in this model has been experimentally justified at the 1\% level 
of accuracy. It is, therefore, important to search for the experimental signatures for the 
SUSY GUT.

It is known that the interaction at  the GUT scale 
($M_{\rm GUT}\sim 10^{16}$GeV) can give 
effects on the lepton flavor violation (LFV) processes,
such as $\mu \rightarrow \e \gamma$, through the LFV slepton 
masses induced by the radiative correction \cite{HKR}.
In particular, the large top-quark Yukawa coupling can be a source 
of sizable LFV in the slepton masses \cite{BH}.  
The precise values of the event rates depend on the detail of the models 
[2-6].
In the minimal SU(5) SUSY GUT the Yukawa interaction of the colored 
Higgs multiplet gives non-negligible LFV masses to only  the 
right-handed sleptons, but not to the left-handed sleptons, and 
the LFV event rates are significantly reduced by destructive interference 
among the diagrams \cite{HMTY}. On the other hand, in 
the SO(10) SUSY GUT two diagrams enhanced by 
$(m_\tau /m_\mu)$ (Fig.~1), which do not interfere with each other,
almost dominate in the LFV event processes since
both the left-handed and the right-handed sleptons can have LFV masses, and 
the branching ratio of $\mu \rightarrow \e \gamma$ 
may reach to the present experimental bound \cite{BHS}.  

It is known that the mass ratios of down-type quarks and charged leptons 
in the first  and the second generations cannot be explained in the minimal SU(5) 
SUSY GUT, while the bottom-tau ratio is
justified in regions $\tan\beta\simeq 2$ or $\gsim 30$ \cite{btau}. One of the 
possible solutions is to introduce higher dimensional operators 
suppressed by the gravitational scale ($M_{\rm G}\sim 10^{18}$GeV).
Since the GUT scale is near to the gravitational scale, it is conceivable that 
the higher dimensional operators may give sizable contribution to the 
fermion masses in the first and the second generations. In  Ref.~\cite{ACH}
the contribution from the higher dimensional operators to the LFV processes
is considered. It is noted that since both the left-handed and the right-handed sleptons 
can have sizable LFV masses at large $\tan\beta$, the branching ratios are 
expected to be enhanced in the general SU(5) SUSY GUT compared with the minimal 
case. In this paper, we evaluate numerically  the 
branching ratio of $\mu \rightarrow \e \gamma$ in the SU(5) SUSY GUT, 
introducing the higher dimensional operators suppressed by the gravitational 
scale. We show that for large $\tan\beta$ it is enhanced by a few orders of 
magnitude compared with that in the minimal case, and that 
the destructive interference among the diagrams disappears.

Let us first discuss the minimal SU(5) SUSY GUT to show that 
the LFV masses for the left-handed sleptons are not induced in this  
case. Both quarks and leptons are embedded in $\phi({\bf 5}^*)$ and 
$\psi({\bf 10})$ as follows,
\begin{eqnarray}
\psi &=& \frac{1}{\sqrt{2}}
\left( \begin{array}{ccccc} 
  0 & \overline{U} & -\overline{U}   & U &  D \\
    &    0         &  \overline{U}   & U &  D \\
    &           &    0               & U &  D \\ 
    &           &                    &  0  &  \overline{E} \\ 
    &           &                    &     & 0  
\end{array} \right), \\
\phi &=& \left(
\begin{array}{ccccc}
\overline{D}&\overline{D}& \overline{D}& E& -N 
\end{array}
\right),
\end{eqnarray}
where $Q(\equiv(U,D))$ and $L(\equiv(N,E))$ are SU(2)$_L$ doublet (left-handed)
quarks and leptons, and $\overline{U}$, $\overline{D}$, and $\overline{E}$ are SU(2)$_L$ 
singlet (right-handed) quarks and leptons. The doublet Higgs supermultiplets $H_f$ and $\overline{H}_f$
in the minimal SUSY standard model (MSSM) are embedded in $H({\bf 5})$ and $\overline{H}({\bf 5}^*)$ with 
the colored Higgs supermultiplets $H_c$ and $\overline{H}_c$, 
\begin{eqnarray}
H &=& \left(
\begin{array}{ccccc}
H_c &H_c & H_c &H^+_f & H_f^0
\end{array}
\right) , 
\nonumber\\
\overline{H}&=& \left( 
\begin{array}{ccccc}
\overline{H}_{c} &\overline{H}_{c}&\overline{H}_{c} &
\overline{H}^-_f & -\overline{H}^0_f
\end{array}
\right).
\end{eqnarray}
In order to break the SU(5) gauge symmetry to those of  the SM , we introduce
an adjoint Higgs supermultiplet $\Sigma(\bf 24)$, whose vacuum expectation value is given as
\begin{equation}
\langle \Sigma \rangle = 
\left( 
\begin{array}{ccccc} 
2&&&&\\  
&2&&&\\  
&&2&&\\  
&&&-3& \\
&&&&-3 
\end{array} 
\right) V .
\end{equation}

The renormalizable superpotential leading to the fermion masses
is given as
\begin{eqnarray}
W_R &=& \frac14 f^{(0)ij}_u \psi^{AB}_i \psi^{CD}_j H^E \epsilon_{ABCDE}
     + \sqrt{2} f^{(0)ij}_d \psi^{AB}_i \phi_{Aj} \overline{H}_B 
\label{Yukawa}
\end{eqnarray}
where  $i$ and $j$ ($=1-3$) are generation indices  and $A$, $B$, $\cdots$ 
($=1-5$) are SU(5) ones. After removing 
unphysical degrees of freedom in the Yukawa coupling constants, they are 
given as
\begin{eqnarray}
f_u^{(0)ij} &=& V_{\rm KM}^{ki} f_{u_k} \e^{i\theta_k} V_{\rm KM}^{kj}, 
\nonumber\\
f_d^{(0)ij} &=& f_{d_i} \delta^{ij}, 
\end{eqnarray}
where $V_{\rm KM}$ corresponds to the Kobayashi-Maskawa (KM) matrix at the GUT
scale and $\theta_i$'s $(i=1-3)$ are additional phases which satisfy
$\theta_1+\theta_2+\theta_3 =0$. After taking a basis of  the SM fields as 
\begin{eqnarray}
\psi_i &\ni& \{Q_i, \e^{-i\theta_j} V_{\rm KM}^{ji\star} \overline{U}_j,\overline{E}_i\}, 
\nonumber\\
\phi_i &\ni& \{\overline{D}_i,L_i\}, \nonumber
\end{eqnarray}
we get 
\begin{eqnarray}
W_R&=&
\phantom{+}f_{d_i} \overline{E}_i L_i \overline{H}_f
  + f_{d_i} Q_i \overline{D}_i \overline{H}_f
  + V_{\rm KM}^{ji} f_{u_j} Q_i \overline{U}_{j} H_f
\nonumber \\
 & &
  + V_{\rm KM}^{ji} f_{u_j} 
    \overline{E}_i \overline{U}_j H_c 
  -  f_{d_i} Q_i L_i \overline{H}_c
\nonumber\\
 & &
  - \frac12 V_{\rm KM}^{ki} f_{u_k} \e^{i\theta_k} V^{kj}_{\rm KM} 
    Q_i Q_j H_c
  + V_{\rm KM} ^{ij\star} f_{d_j} \e^{-i\theta_i}  
    \overline{U}_i \overline{D}_j \overline{H}_c. 
\label{minimal}
\end{eqnarray}
Terms in the first line represent the superpotential of the MSSM.
In the fourth term of the right-handed side the right-handed leptons have 
flavor-violating interaction with the colored Higgs multiplet, which is 
controlled by the KM matrix. This
interaction leads to  non-negligible LFV masses for
the right-handed sleptons through the radiative correction at one-loop level
as,
\begin{eqnarray}
[m^2_{\bar{e}}]^i_j &=& 
-\frac{3}{8\pi^2} 
f_{u_3}^2 V_{\rm KM}^{3i} V_{\rm KM}^{3j\star}
(3 +a_0^2) m_0^2
\log\frac{M_{\rm G}}{M_{\rm GUT}},
\end{eqnarray}
with $i\ne j$. In this paper we assume the minimal supergravity scenario 
for keeping the universality of SUSY breaking parameters at tree level, 
and $m_0$ and $a_0$ are the SUSY breaking scalar mass and the trilinear scalar
coupling parameter at the tree level \cite{nilles}.
Here, we ignored the up-type Yukawa coupling constants except for
that of top quark. On the other 
hand, the Yukawa coupling of the left-handed sleptons to 
the colored Higgs
multiplet is diagonal. Though the off-diagonal components are induced at 
the higher orders, the effect on the LFV left-handed slepton masses is negligibly 
small.

Both the Yukawa coupling constants for down-type quarks and those for leptons are 
given by $f_{d_i}$ at the GUT scale in the minimal
case. As explained in introduction, however, this is 
not justified at least for the first and the second generations. Since the result that 
only the right-handed sleptons have LFV masses depends on this unrealistic
assumption for the Yukawa coupling, we investigate the LFV processes 
taking into account higher dimensional operators with $\Sigma$ in the 
superpotential.

We consider the following superpotential including higher dimensional operators
up to the dimension five,
\begin{eqnarray}
W &=& \phantom{+} W_R\nonumber \\
& & +\frac1{4 M_{\rm G}} f^{(1)ij}_u 
\left[ \psi^{AB}_i \psi^{CD}_j 
      +\psi^{AB}_j \psi^{CD}_i \right] 
\Sigma^E_A H^F \epsilon_{BCDEF}  \nonumber \\
& & + \frac1{4 M_{\rm G}} f^{(2)ij}_u 
\left[ \psi^{AB}_i \psi^{CD}_j 
      -\psi^{AB}_j \psi^{CD}_i \right] 
\Sigma^E_A H^F \epsilon_{BCDEF}  \nonumber \\
& & + \frac{\sqrt{2}}{M_{\rm G}} f^{(1)ij}_d 
\left[ \Sigma^A_B \psi^{BC}_i 
      +\Sigma^C_B \psi^{AB}_i \right] 
\phi_{Aj} \overline{H}_C             \nonumber \\
& & + \frac{\sqrt{2}}{M_{\rm G}} f^{(2)ij}_d 
\left[ \Sigma^A_B \psi^{BC}_i 
      -\Sigma^C_B \psi^{AB}_i \right] 
\phi_{Aj} \overline{H}_C,
\end{eqnarray}
where $W_R$ is given by Eq.~(\ref{Yukawa}).
In order to see explicitly the interaction of the colored Higgs multiplets after 
inserting the vacuum expectation value into $\Sigma$, we decompose $W$ into 
\begin{eqnarray}
W&=& 
\phantom{+} f_l^{ij} \overline{E}_i L_j \overline{H}_f
  + f_d^{ij} Q_i  \overline{D}_j \overline{H}_f
  + f_u^{ij} Q_i  \overline{U}_j H_f  \nonumber 
\nonumber\\
 & &
  + f_{c_R}^{ij} 
    \overline{E}_i \overline{U}_j H_c 
  + f_{c_L}^{ij} Q_i Q_j H_c
\nonumber\\
 & &
  + f_{\overline{c}_R}^{ij} 
    \overline{U}_i \overline{D}_j \overline{H}_c
  + f_{\overline{c}_L}^{ij} Q_i L_j \overline{H}_c.
\end{eqnarray}
Here, the Yukawa coupling constants of the doublet Higgs multiplets are given as
\begin{eqnarray}
f_l^{ij}&=&f_d^{(0)ij}-\frac{6V}{M_{\rm G}}f_d^{(1)ij}, 
\nonumber\\
f_d^{ij}&=&f_d^{(0)ij}-\frac{V}{M_{\rm G}}f_d^{(1)ij}+\frac{5V}{M_{\rm G}}f_d^{(2)ij},  
\nonumber\\
f_u^{ij}&=&f_u^{(0)ij}-\frac{3V}{2M_{\rm G}}f_u^{(1)ij}+\frac{5V}{2M_{\rm G}}f_u^{(2)ij}, 
\label{yukawa_mass}
\end{eqnarray}
and those of the colored Higgs multiplets are
\begin{eqnarray}
f_{c_R}^{ij}
&=& f_u^{(0)ij}+\frac{V}{M_{\rm G}}f_u^{(1)ij}+\frac{5V}{M_{\rm G}}f_u^{(2)ij}, 
\nonumber \\
f_{\overline{c}_L}^{ij}
&=& -f_d^{(0)ij}+\frac{V}{M_{\rm G}}f_d^{(1)ij}+\frac{5V}{M_{\rm G}}f_d^{(2)ij}, 
\nonumber \\
f_{\overline{c}_R}^{ij}
&=& f_d^{(0)ij}+\frac{4V}{M_{\rm G}}f_d^{(1)ij}, 
\nonumber \\
 f_{c_L}^{ij} 
&=& -\frac12 f_u^{(0)ij}-\frac{V}{2M_{\rm G}}f_u^{(1)ij}. 
\end{eqnarray}

The contribution from the higher dimensional 
operators to the Yukawa coupling constants of the doublet Higgs multiplets can be  
as large as ${V}/M_{\rm G} \sim 10^{-2}$ without violation of unitarity. Then, from 
Eq.~(\ref{yukawa_mass}) the fermion masses for the first and the second generations
are determined by both the  renormalizable and the higher dimensional terms, while 
those of the third generation are almost determined by the renormalizable terms.
In this case, the Yukawa coupling constants of the colored Higgs multiplet to 
the right-handed leptons are not necessarily diagonalized by the KM matrix 
unlike in the minimal case given in Eq.~(\ref{minimal}). More remarkablely, 
we can not diagonalize both the Yukawa coupling constants of 
the left-handed leptons to the doublet and the colored Higgs multiplets simultaneously
unless both $f_d^{(1)}$ and $f_d^{(2)}$ are vanishing. This means
that the left-handed sleptons may also receive LFV masses from the radiative 
correction.

Let us evaluate the LFV event rate of $\mu \rightarrow \e \gamma$. 
Since degrees of freedom in the higher dimensional operators are huge, 
we parameterize the Yukawa couplings as follows,
\begin{eqnarray}
f_l^{ij}&=& f_{l_i} \delta^{ij}, 
\nonumber\\
f_d^{ij}&=& f_{d_i} \delta^{ij}, 
\nonumber\\
f_u^{ij}&=& V_{\rm KM}^{ji} f_{u_j}, 
\end{eqnarray}
and 
\begin{eqnarray}
f_{c_R}^{ij}
&=&
V_{\bar{e}}^{ki} f_{c_{Rk}} V_{\bar{u}1}^{kj}, 
\nonumber \\
 f_{\overline{c}_L}^{ij} 
&=& 
V_{q1}^{ki} f_{\overline{c}_{Lk}} V_{l}^{kj}, 
\nonumber \\
f_{\overline{c}_R}^{ij}
&=& 
V_{\bar{u}2}^{ki} f_{\overline{c}_{Rk}} V_{\bar{d}}^{kj}, 
\nonumber \\
 f_{c_L}^{ij}
&=& 
V_{q2}^{ki} f_{c_{Lk}} V_{q2}^{kj}.
\end{eqnarray}

The LFV SUSY breaking parameters of slepton
are given at one-loop level keeping the logarithmic terms as
\begin{eqnarray}
[m^2_{\bar{e}}]^i_j &=& 
-\frac{3}{8\pi^2} 
f_{c_{R3}}^2 V_{\bar{e}}^{3i} V_{\bar{e}}^{3j\star}
(3 +a_0^2) m_0^2
\log\frac{M_{\rm G}}{M_{\rm GUT}},
\label{right}
\end{eqnarray}
\begin{eqnarray}
[m^2_{l}]^i_j &=& 
-\frac{3}{8\pi^2} 
f_{\overline{c}_{L3}}^2 V_{l}^{3i} V_{l}^{3j\star}
(3  +a_0^2) m_0^2 
\log\frac{M_{\rm G}}{M_{\rm GUT}},
\label{left}
\end{eqnarray}
\begin{eqnarray}
A_l^{ij} &=& 
-\frac{9}{16\pi^2} 
\left(
f_{\overline{c}_{L3}}^2 V_{l}^{3i} V_{l}^{3j\star} f_l^j
+
f_l^i f_{c_{R3}}^2 V_{\bar{e}}^{3i} V_{\bar{e}}^{3j\star}
\right)a_0 m_0
\log\frac{M_{\rm G}}{M_{\rm GUT}},
\label{agut}
\end{eqnarray}
with $ i \ne j$. 
Here, they are defined as
\begin{eqnarray}
-{\cal L}_{\rm SB} &=& [m^2_{\bar{e}}]_i^j 
\tilde{\bar{e}}^{i \dagger} \tilde{\bar{e}}_{j} 
+[m^2_{l}]_i^j 
\tilde{l}^{i \dagger} \tilde{l}_{j} 
+(A_l^{ij} \tilde{\bar{e}}_{i} \tilde{l}_j \bar{h}_f+ h.c.)
\end{eqnarray} 
where $\tilde{\bar{e}}$ and $\tilde{l}$ are the right-handed and the left-handed sleptons, 
and $\bar{h}_f$ is one of the doublet Higgs bosons. 
Also, we assume the 
hierarchical structure of $f_{c_{Rk}}$ and  $f_{\overline{c}_{Lk}}$.
From the above equations, the LFV masses depend on the magnitude of 
$f_{c_{R3}}$ and $f_{\overline{c}_{L3}}$, and the mixing matrixes 
$V_{\bar{e}}$ and $V_{l}$.
Notice that when $\tan\beta$ is large, the off-diagonal terms in
the left-handed slepton mass matrix can be  as large as 
those of the right-handed sleptons provided that $V_l^{3i}V_l^{3j\star}\sim 
V_{\bar e}^{3i}V_{\bar e}^{3j\star}$.
In such a case the branching ratio of $\mu \rightarrow \e \gamma$ 
is 
almost determined by two diagrams (Fig.~1), which are proportional to 
the mass of tau lepton. They do not generally interfere with each other.
If only the right-handed sleptons have the sizable LFV 
masses, every diagram contributing to  
$\mu \rightarrow \e \gamma$ is proportional to the muon mass and
destructive interference tends to occur. Therefore, while at small  $\tan\beta$ the 
effect from the higher dimensional operators is small, at large 
$\tan\beta$ the destructive interference disappears and the branching 
ratio can be larger than in the minimal case.

In Fig.~2,
we show dependence of the branching ratio of $\mu\rightarrow \e \gamma$ on 
the right-handed selectron mass ($m_{\tilde{\e}_R}$) for $\tan\beta$=6 and 30. As an illustration, 
we assume 
$V_{\bar{e}}$= $V_{l}$ = $V_{\rm \rm KM}$, 
$f_{l_3} = f_{\overline{c}_{R3}}  = -f_{\overline{c}_{L3}}$,
and  $f_{u_3} = f_{c_{R3}}= -f_{c_{L3}}/2$ at the GUT scale. 
In this and next figures, the lines denoted as "nonminimal" are 
calculated on this assumption. This is 
realized, for example, at $\tan\beta=30$ by taking the following
choice, 
\begin{eqnarray}
    f_u^{(0)} &=&  \left(
    \begin{array}{rrc}
      9.8 \times 10^{-5}   & -4.5 \times 10^{-4}  &  3.2 \times 10^{-3} \\
     -4.5 \times 10^{-4}   &  2.3 \times 10^{-3}  & -2.3 \times 10^{-2} \\
      3.2 \times 10^{-3}   & -2.3 \times 10^{-2}  &  0.65               \\
    \end{array}
    \right),
\\
    f_d^{(0)} &=&  \left(
    \begin{array}{rrc}
      1.6 \times 10^{-4}   &  8.3 \times 10^{-6}  &  1.1 \times 10^{-7} \\
     -1.8 \times 10^{-3}   &  7.7 \times 10^{-3}  &  2.8 \times 10^{-4} \\
      6.9 \times 10^{-4}   & -4.8 \times 10^{-3}  &  0.21               \\
    \end{array}
    \right),
\\
    f_d^{(1)} &=&  \left(
    \begin{array}{rcc}
      1.6 \times 10^{-3}   &  1.4 \times 10^{-4}  &  1.8 \times 10^{-6} \\
     -2.9 \times 10^{-2}   & -0.09                &  4.6 \times 10^{-3} \\
      1.2 \times 10^{-2}   & -8.1 \times 10^{-2}  & -0.39                \\
    \end{array}
    \right),
\\
    f_d^{(2)} &=&  \left(
    \begin{array}{rrc}
      1.6 \times 10^{-3}   & -1.4 \times 10^{-4}  & -1.8 \times 10^{-6} \\
      2.9 \times 10^{-2}   & -8.4 \times 10^{-2}  & -4.6 \times 10^{-3} \\
     -1.2 \times 10^{-2}   &  8.1 \times 10^{-2}  & -0.39               \\
    \end{array}
    \right),
\end{eqnarray}
and $f_u^{(1)}$ and $f_u^{(2)}$ are zero. Also, we show the 
branching ratio in the minimal case where $V_{\bar{e}}$ is $V_{\rm KM}$ and 
$V_{l}$ is the unit matrix. 
In this figure we take the bino mass 60GeV, $a_0$ =0, the higgsino mass positive, and the top quark mass 175GeV.
In our actual calculation, we solve the renormalization group equations of the diagonal
elements in the sfermion mass matrixes and use 
Eqs.~(\ref{right},\ref{left},\ref{agut}) for the off-diagonal components in the slepton mass matrixes.
Also, we impose the radiative
breaking condition of the SU(2)$_L\times$U(1)$_Y$ symmetry, and the experimental
constraints including the anomalous magnetic dipole moment of muon. 
The detailed formula of $\mu\rightarrow \e \gamma$ is given in 
Ref.~\cite{HMTYfull}. 
We also show the experimental upper bound given in Ref.~\cite{mueg}.
For $\tan\beta=30$ the destructive interference at 
$m_{\tilde{\e}_R}\simeq 300$GeV in the minimal case disappears in the 
nonminimal case, 
and  the branching ratio is enhanced by two orders of magnitude for 
$m_{\tilde{\e}_R} \gsim 400$GeV. On the other hand, for $\tan\beta=6$ 
the branching ratio is almost the same as in the minimal case.

In Fig.~3 the branching ratio of $\mu\rightarrow \e \gamma$ is shown 
as a function of $\tan\beta$ for the minimal and the nonminimal case. 
Here, $m_{\tilde{\e}_R}= 200$GeV  and 300GeV and 
the other input parameters are taken as same as in Fig.~2.  
At large $\tan\beta$ the branching ratio is enhanced in the nonminimal
case, and it can reach to one order of magnitude below the experimental bound.
Notice that the branching ratio is proportional to 
$|V_{\bar{e}}^{33} V_{\bar{e}}^{32\star} V_{l}^{33} V_{l}^{31\star}|^2$
or 
$|V_{\bar{e}}^{33} V_{\bar{e}}^{31\star} V_{l}^{33} V_{l}^{32\star}|^2$
at large $\tan\beta$. For example, if the mixing angles
$V_{l}^{31}$ or $V_{l}^{32}$ is four times larger than those
in the KM matrix, a part of the parameter space is already excluded by 
the experimental bound.

Comments on other processes which may be important for large $\tan\beta$ 
are in order. First,
the EDM of electron may be induced at one-loop level if both the left-handed
and the right-handed sleptons have LFV masses. It is proportional to square root 
of the branching ratio of $\mu \rightarrow \e \gamma$ at large $\tan\beta$ as 
is shown in Ref.~\cite{ACH}, 
\begin{equation}
\left|\frac{d_\e}{e} \right|
= 3.5\times 10^{-26} {\rm c.m.} \times
\frac{
\sqrt{2}{\rm Im} 
V_{\bar{e}}^{31} V_{\bar{e}}^{33\star}
V_{l}^{31} V_{l}^{33\star}
}
{
\sqrt{
\left|V_{\bar{e}}^{31} V_{\bar{e}}^{33\star}\right|^2
+\left|V_{l}^{31} V_{l}^{33\star}\right|^2
}}
\left(\frac{{\rm Br}(\mu\rightarrow \e \gamma)}{4.9 \times10^{-11}} \right)^{\frac12}.
\end{equation}
This depends on the phase and the mixing matrixes, and the EDM of electron can be 
as large as the experimental upper bound
$|d_\e/e| \le 4\times10^{-27}{\rm c.m.}$ \cite{edm}. Next, the amplitude of 
$b\rightarrow s\gamma$ transition is enhanced when $\tan\beta$ is large.
In the SM the $b \rightarrow s \gamma$ process is induced by an effective operator 
$ \bar{s}_L i\sigma_{\mu\nu} F^{\mu\nu} b_R$.  In the MSSM 
there are contributions to the operator from the loop of charged Higgs boson and top quark 
and the loop of chargino and stop. It is known that the latter contribution
interferes constructively or destructively with the SM contribution,
depending on the SUSY parameters.
Moreover, in the SUSY GUTs there are additional
contributions to the above operator as well as another operator 
$\bar{s}_R i\sigma_{\mu\nu} F^{\mu\nu} b_L$. Since these contributions
depend on unknown mixing parameters at the GUT scale, in above analysis
we do not impose  the $b\rightarrow s \gamma$ constraint although we have 
checked at least the MSSM contributions do not exceed the experimental
upper bound.  Finally, the exchange of the colored Higgs multiplets
induces instability of proton, and no observation of proton decay  gives
the strongest constraint \cite{proton} on 
\begin{eqnarray}
\frac{f_{c_R}^{ij} f_{\overline{c}_R}^{kl}}{M_{H_c}^{eff}}
\overline{E}_i \overline{U}_j \overline{U}_k \overline{D}_l,&
&\frac{f_{c_L}^{ij} f_{\overline{c}_L}^{kl}}{M_{H_c}^{eff}}
Q_i Q_j Q_k L_l.
\end{eqnarray}
In this analysis, we do not include the constraint from proton
decay, since it is sensitive to the precise value of the mass
parameter $M^{eff}_{H_c}$ which can be different from the colored
Higgs mass in some models \cite{HMY}.

In summary, we have evaluated the branching ratio of $\mu \rightarrow \e \gamma$ 
in  the supersymmetric SU(5) grand unified theory taking account of the higher 
dimensional operators, and have shown that it reaches to one order of magnitude 
below the experimental bound for $\tan\beta>30$ even if 
$V_{\bar{e}}$= $V_{l}$ = $V_{\rm KM}$. If the mixing angles
are larger, the branching ratio is further enhanced. 
 
\newpage
%
%
\newcommand{\Journal}[4]{{\sl #1} {\bf #2} {(#3)} {#4}}
\newcommand{\APJ}{Ap. J.}
\newcommand{\CJP}{Can. J. Phys.}
\newcommand{\MPL}{Mod. Phys. Lett.}
\newcommand{\NC}{Nuovo Cimento}
\newcommand{\NP}{Nucl. Phys.}
\newcommand{\PL}{Phys. Lett.}
\newcommand{\PR}{Phys. Rev.}
\newcommand{\PRep}{Phys. Rep.}
\newcommand{\PRL}{Phys. Rev. Lett.}
\newcommand{\PTP}{Prog. Theor. Phys.}
\newcommand{\SJNP}{Sov. J. Nucl. Phys.}
\newcommand{\ZP}{Z. Phys.}

\newpage
%
%
%
%
\begin{figure}[p]
\epsfxsize=15cm
\caption
{
Feynman diagrams contributing to $\mu\rightarrow \e \gamma$, which 
are proportional to $m_\tau$. In these diagrams, $\tilde{\e}_{L(R)}$, 
$\tilde{\mu}_{L(R)}$, and $\tilde{\tau}_{L(R)}$  are the left-handed 
(right-handed) selectron, smuon,and stau respectively, and $\tilde{B}$ is bino.
The symbol $\mu$ refers to the higgsino mass. The arrows represent
the chirality. 
}
\end{figure}
%
%
%
%
\begin{figure}[p]
\epsfxsize=15cm
\caption
{
Dependence of the branching ratio of $\mu\rightarrow \e \gamma$
on the right-handed selectron mass in our model for $\tan\beta=6$
(dashed lines) and 30 (solid lines). The thick lines are for 
the nonminimal case in which  $V_{\bar{e}}$ and $V_{l}$ are the same as 
$V_{\rm KM}$, and the thin lines are for the minimal case in which 
$V_{\bar{e}}=V_{\rm KM}$ and $V_{l}={\bf 1}$. In this figure we 
choose the bino mass 60GeV, $a_0$ =0, the higgsino mass positive, and 
the top quark mass 175GeV. 
The long-dashed line is the 
experimental upper bound.
}
\end{figure}
%
%
\begin{figure}
\epsfxsize=15cm
\caption
{
Dependence of the branching ratio of $\mu\rightarrow \e \gamma$
on $\tan\beta$ for  the right-handed selectron mass 200GeV 
(dashed lines) and 300GeV (solid lines). The thick lines are for 
the nonminimal case that $V_{\bar{e}}$ and $V_{l}$ are the same as 
$V_{\rm KM}$, and the thin lines are for the minimal case in which 
$V_{\bar{e}}=V_{\rm KM}$ and $V_{l}={\bf 1}$. In this figure we 
choose the bino mass 60GeV, $a_0$ =0, the higgsino mass positive. 
The long-dashed line is the experimental upper bound.
}
\end{figure}
%
%
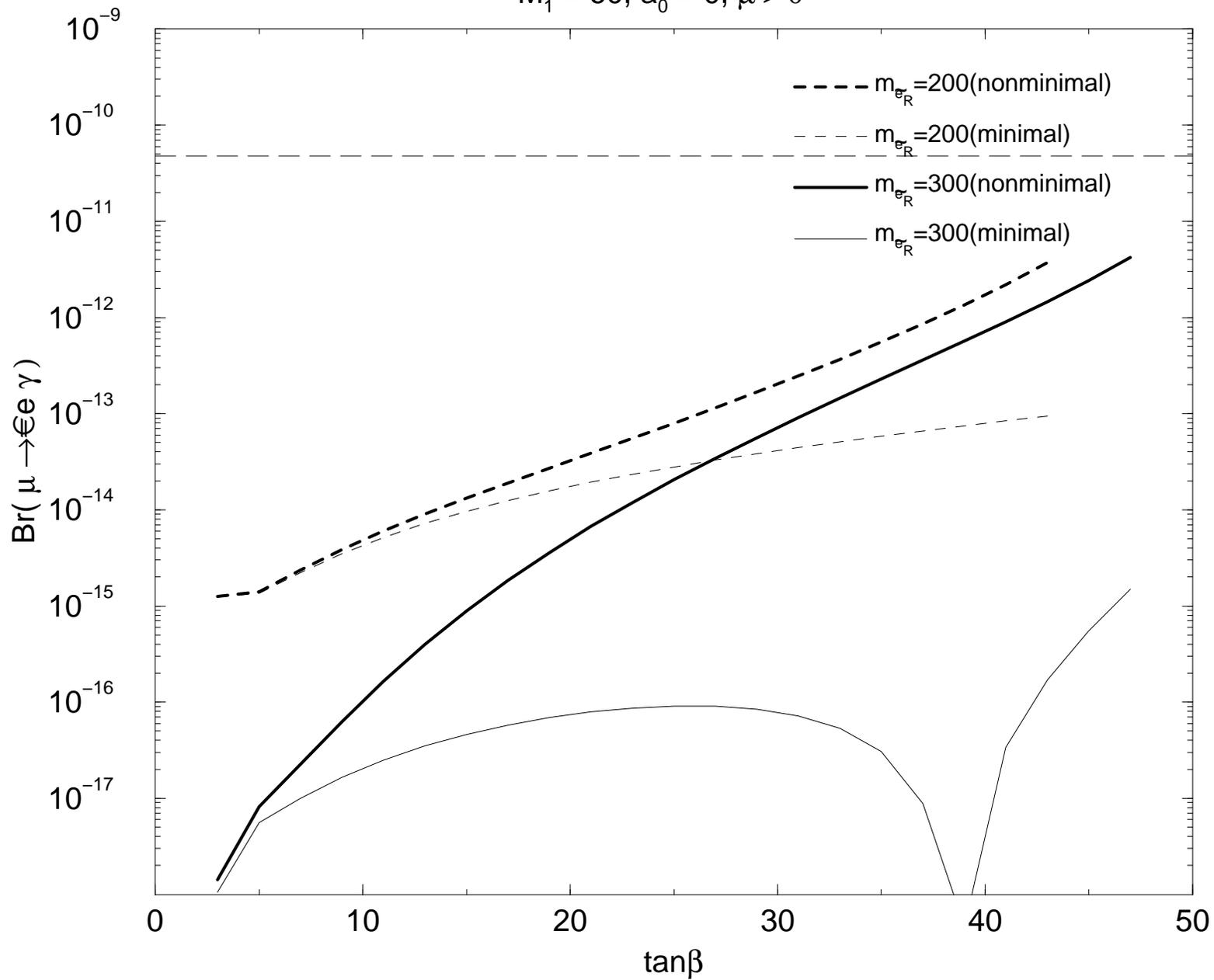
\end{document}